\documentclass{elsart}
\usepackage{amssymb}
\journal{Physics Letters A}

\usepackage{graphicx}
\usepackage{dcolumn}
\usepackage{bm}
\usepackage{psfrag}
\usepackage{subfigure}

\newcommand{\be}{\begin{eqnarray}}
\newcommand{\ee}{\end{eqnarray}}
\newcommand{\bn}{\begin{eqnarray*}}
\newcommand{\en}{\end{eqnarray*}}

\newcommand{\nn}{\nonumber \\}
\newcommand{\nl}{\\}

\renewcommand{\vec}[1]{\mbox{\boldmath$#1$}}
\renewcommand{\d}{\mbox{\rm d}}

\newcommand{\lmbar}{\mathchar'26\mkern-9mu \lm}

\newcommand{\al}{\ensuremath{\alpha}}
\newcommand{\bt}{\ensuremath{\beta}}
\newcommand{\sg}{\ensuremath{\sigma}}

\newcommand{\dl}{\ensuremath{\delta}}
\newcommand{\lm}{\ensuremath{\lambda}}

\newcommand{\Dl}{\ensuremath{\Delta}}

\newcommand{\Gm}{\ensuremath{\Gamma}}

\newcommand{\pvec}{\ensuremath{\vec{p}}}
\newcommand{\Pvec}{\ensuremath{\vec{P}}}

\newcommand{\xvec}{\ensuremath{\vec{x}}}

\newcommand{\Xvec}{\ensuremath{\vec{X}}}

\newcommand{\xivec}{\ensuremath{\vec{\xi}}}

\newcommand{\nabvec}{\ensuremath{\vec{\nabla}}}

\newcommand{\hb}{\ensuremath{\hbar}}

\newcommand{\lt}{\ensuremath{\left}}
\newcommand{\rt}{\ensuremath{\right}}

\renewcommand{\d}{\mbox{\rm d}}

\newcommand{\Lie}{{\cal L}_{\xivec} \Xvec}

\begin{document}

\begin{frontmatter}



\title{A New Perspective on Path Integral Quantum Mechanics in Curved Space-Time}%


\author[label1,label2]{Dinesh Singh\corauthref{cor}},
\corauth[cor]{Corresponding author.}
\ead{dinesh.singh@uregina.ca}
\author[label1]{Nader Mobed}
\ead{nader.mobed@uregina.ca}
\address[label1]{Department of Physics, University of Regina \\
Regina, Saskatchewan, S4S 0A2, Canada}
\address[label2]{Department of Physics and Engineering Physics, University of Saskatchewan \\
Saskatoon, Saskatchewan, S7N 5E2, Canada}

\begin{abstract}

A fundamentally different approach to path integral quantum mechanics in curved space-time
is presented, as compared to the standard approaches currently available in the literature.
Within the context of scalar particle propagation in a locally curved background,
such as described by Fermi or Riemann normal co-ordinates, this approach requires use of
a constructed operator to rotate the initial, intermediate, and final
position ket vectors onto their respective local tangent spaces, defined at each local time step
along some arbitrary classical reference worldline.
Local time translation is described using a quantum mechanical representation of
Lie transport, that while strictly non-unitary in operator form, nevertheless correctly
recovers the free-particle Lagrangian in curved space-time, along with new contributions.
This propagator yields the prediction that all probability violating terms due to curvature
contribute to a quantum violation of the weak equivalence principle, while the remaining
terms that conserve probability also correspondingly satisfy the weak equivalence principle,
at least to leading-order in the particle's Compton wavelength.
Furthermore, this propagator possesses an overall curvature-dependent and gauge-invariant phase factor that
can be interpreted as the gravitational Aharonov-Bohm effect and Berry's phase.

\end{abstract}

\begin{keyword}
path integrals \sep quantum mechanics \sep curved space-time \sep weak equivalence principle violation
\PACS 04.60.Gw \sep 03.65.-w \sep 03.65.Ca \sep 03.65.Sq
\end{keyword}
\end{frontmatter}

\section{Introduction}

It is an undeniable fact that the Feynman path integral approach to quantum mechanics \cite{Feynman,Feynman-Hibbs}
has made significant contributions towards the present-day understanding of theoretical physics,
especially in the study of subatomic particle theory, condensed matter, and statistical mechanics \cite{Kleinert}.
While as a computational tool within quantum mechanics it is arguably less relevant for physical problems that are much
easier to solve via the standard canonical quantization approach, its inherent value comes from the physical
insight it provides on the relationship between classical and quantum phenomena within a unified mathematical
framework \cite{Sakurai}.
In terms of a conceptually intuitive view of quantum mechanics as a ``sum-over-histories'' for all possible
{\em classical worldlines} in space-time, the path integral approach is well-suited for better
understanding quantum phenomena in the presence of a classical gravitational background
\cite{Bastianelli-van-Nieuwenhuizen}, which may lay the groundwork towards finding
a self-consistent theory of quantum gravity.

At present, this ultimate goal has remained elusive for over 70 years, in no small part due to the lack of
any physical data to provide grounding and direction for building the theory.
Nonetheless, it can be argued that many useful theoretically-motivated efforts already exist,
including the path integral approach.
This is most evident through the pioneering efforts of DeWitt for developing much of the mathematical machinery now used
for this avenue of searching for quantum gravity \cite{DeWitt}.
Within a more modest semiclassical context, such as quantum field theory in curved space-time, many advances
have also taken place from applying a path integral approach---which avoids certain conceptual issues related
to the loss of a global Poincar\'{e} symmetry used to denote the inherent nature of particles---while recovering
interesting physical consequences originally obtained by canonical methods.

Notwithstanding the standard approaches to path integral quantum gravity or quantum field theory
in curved space-time \cite{Kleinert,Kleinert-1,Parker-Toms,Parker-pi,Bastianelli,Ferraro}, there are a number of relevant
conceptual questions to raise about the precise physical nature of quantum particle propagation in a non-trivial
gravitational background.
For example, to what extent does the curved space-time manifold have a mathematically smooth structure when
applying the path integral approach?
Is it possible to effectively perform the sum-over-histories when the intermediate classical paths deviate
significantly from a classical geodesic?
Is the presence of space-time torsion necessarily required to correctly define the path integral in curved
space-time \cite{Kleinert,Kleinert-1}?
Are the intermediate classical paths required to preserve local causality or should they be free to trace out
causality-violating worldlines, including ones that imply propagation into the local past \cite{Ferraro}?
To what extent can a coarse-grained skeletonized form of the path integral in curved space-time be identified with
a truly continuum form in the limit as the finitely chosen time step becomes infinitesimally small?
In the absence of physical data, it seems unclear whether any of these questions can be readily answered.

Instead of addressing these questions outright, the approach taken in this Letter is to consider the idea
of path integral quantum mechanics for a scalar particle from a fundamentally different perspective,
while simultaneously preserving Feynman's original vision to the best extent possible.
This is done by formulating the problem in terms of either Fermi or Riemann normal co-ordinates and
employing the orthonormal tetrad formalism \cite{Poisson} to relate all quantities with respect to local tangent
spaces at successive local time steps.
As a result of taking this approach, a number of interesting physical predictions emerge from the scalar
particle propagator that are not readily evident in the more standard approaches found in the literature---while
still preserving its expected properties in flat space-time.
Potentially observable implications also appear for future consideration.
Though full computational details of this approach will be forthcoming in a much longer
publication \cite{Singh-Mobed-prep-1}, it is possible to still outline some of the essential features
of this approach.
All computations are performed assuming the curvature conventions of Misner, Thorne, and Wheeler \cite{MTW}.
As well, $c=1$ is assumed throughout.

\begin{figure}
\psfrag{tau}[cc][][1.8][0]{\Large $\tau$}
\psfrag{x}[cc][][1.8][0]{\Large $\xvec$}
\psfrag{T}[cc][][1.8][0]{\Large $T$}
\psfrag{X}[cc][][1.8][0]{\Large $\Xvec$}
\psfrag{xi}[cc][][1.8][0]{\Large $\xivec$}
\psfrag{U-proj}[cc][][1.8][0]{\Large $U_{\rm Proj.}(\tau, \xvec)$}
\psfrag{x-ket}[cc][][1.8][0]{\Large $|x^\mu\rangle$}
\psfrag{X-ket}[cc][][1.8][0]{\Large $|\Xvec_{\rm (G)}^{\hat{\mu}} (\tau, \xvec) \rangle$}
\hspace{1.0cm}
\begin{minipage}[t]{0.3 \textwidth}
\centering
{
\rotatebox{0}{\includegraphics[width = 8.5cm, height = 6.5cm, scale = 1]{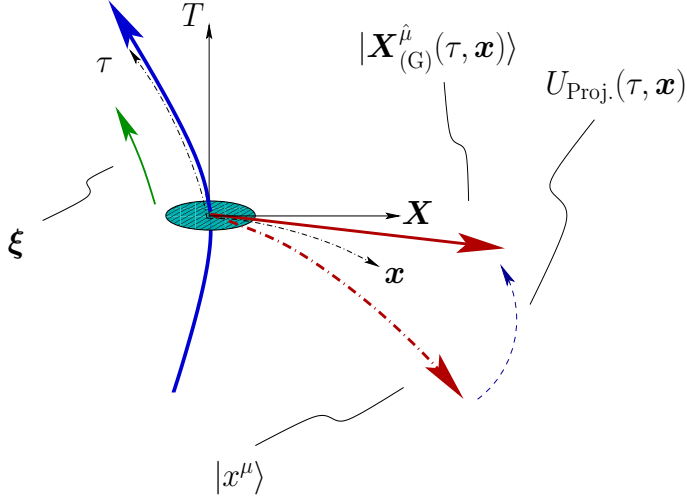}}}
\end{minipage}%
\caption{\label{fig:ket-rotation} A projection operator $U_{\rm Proj.}(\tau, \xvec)$ is used to transform the
position ket vector $\lt| x^\mu \rt\rangle$, defined with respect to Fermi or Riemann normal co-ordinates,
into $|\Xvec_{\rm (G)}^{\hat{\mu}} (\tau, \xvec) \rangle = U_{\rm Proj.}(\tau, \xvec) \lt| x^\mu \rt\rangle$
that exists on a local tangent space defined at $\tau$.
Proper time translation of $|\Xvec_{\rm (G)}^{\hat{\mu}} (\tau, \xvec) \rangle$ is determined according to Lie transport
with respect to a vector field $\xivec$ that is tangent to the reference worldline.}
\end{figure}

\section{Formalism}

The key behind this new perspective follows from identifying a classical geometric description
of position in Fermi or Riemann normal co-ordinates with an equivalent representation of a corresponding ket vector
in Hilbert space.
Suppose that $x^\mu = (\tau, \xvec(\tau))$ refers to the normal co-ordinate system and
$\bar{e}^{\hat{\mu}}{}_\nu = \dl^\mu{}_\nu + \tilde{R}^\mu{}_\nu$ is the orthonormal tetrad, such that hatted indices
refer to a local Lorentz frame coinciding with the tangent space, and $\tilde{R}^\mu{}_\nu$ is a two-indexed
space-time curvature dependent contribution, not to be confused with the Ricci tensor.
Given that $g_{\mu \nu} = \eta_{\hat{\al} \hat{\bt}} \, \bar{e}^{\hat{\al}}{}_\mu \, \bar{e}^{\hat{\bt}}{}_\nu
\approx \eta_{\mu \nu} + 2 \tilde{R}_{(\mu \nu)}$,
it is straightforward to determine that \cite{Poisson-1}
\be
{}^F \tilde{R}^\mu{}_\nu & = & \lt[{1 \over 2} \, {}^F R^\mu{}_{lm0} (\tau) \, \dl^0{}_\nu
+ {1 \over 6} \, {}^F R^\mu{}_{lmk} (\tau) \, \dl^k{}_\nu \rt] \dl x^l \, \dl x^m
\nn
\label{FNC-curvature}
\ee
in Fermi normal co-ordinates, while
\be
{}^R \tilde{R}^\mu{}_\nu & = & {1 \over 6} \, {}^R R^\mu{}_{\al \bt \nu} (\tau) \, \dl x^\al \, \dl x^\bt
\label{RNC-curvature}
\ee
is its counterpart in Riemann normal co-ordinates, where $R^\mu{}_{\al \bt \nu} (\tau)$ is the local expression for
the Riemann tensor, and $\dl x^\mu$ is interpreted as a space-time quantum fluctuation
with $|\dl x^\mu| \ll |x^\mu|$ to ensure that $\tilde{R}^\mu{}_\nu \ll \dl^\mu{}_\nu$.

Suppose now that $\lt| x^\mu \rt\rangle = \lt| (\tau, \xvec) \rt\rangle$ defines the position ket vector for normal
co-ordinates.
Then, from noting that $\Xvec_{\rm (G)}^{\hat{\mu}} (\tau, \xvec) = \bar{e}^{\hat{\mu}}{}_\nu \, x^\nu$
with $\eta_{\hat{\al} \hat{\bt}} \, \Xvec_{\rm (G)}^{\hat{\al}} \, \Xvec_{\rm (G)}^{\hat{\bt}} = g_{\mu \nu} \,
x^\mu \, x^\nu$, it is possible to identify an expression \cite{Sakurai} for the position ket
vector in a local Lorentz frame as $|\Xvec_{\rm (G)}^{\hat{\mu}} (\tau, \xvec) \rangle =
\lt| \bar{e}^{\hat{\mu}}{}_\nu \, x^\nu \rt\rangle = U_{\rm Proj.}(\tau, \xvec) \lt| x^\mu \rt\rangle$, where
\be
U_{\rm Proj.}(\tau, \xvec) & = & 1 + \tilde{R}^\bt{}_\al \, \xvec^\al \, \nabvec_\bt
\ = \
1 + {i \over \hb} \, \tilde{R}_{\bt \al} \, \xvec^\al \, \pvec^\bt
\label{U-Proj.}
\ee
is an operator to project local space-time curvature onto a local tangent space,
in terms of position and canonical momentum operators $\xvec^\al$ and $\pvec^\al$.
By transforming the position ket vector defined within the curved space-time background into
an equivalent expression within the local Lorentz frame, it is possible to incorporate local gravitational
effects ``G'' within a standard Hilbert space setting in flat space-time for each instance of $\tau$,
thereby allowing the standard tools of Feynman path integration to become available for future use.
A schematic representation of this operation is displayed in Figure~\ref{fig:ket-rotation}.

The next step is to determine the local time translation operator when applied to $|\Xvec_{\rm (G)}^{\hat{\mu}}
(\tau, \xvec) \rangle$ on the local tangent space.
By analogy with the geometric description of Lie transport involving $\tau \rightarrow \tau + \Dl \tau$ \cite{Poisson},
the transformed position ket vector is denoted by
\be
\lefteqn{
\lt| \Xvec_{\rm (G)}^{'\hat{\mu}} \lt(\tau + \Dl \tau, \xvec + \Dl \xvec\rt) \rt\rangle
\ = \
\lt| \Xvec_{\rm (G)}^{\hat{\mu}} \lt(\tau + \Dl \tau, \xvec + \Dl \xvec\rt) - \Dl \tau \, [(\Lie)^{\hat{\mu}}
\lt(\tau, \xvec\rt)] \rt\rangle
}
\nn
&&{} = \ \lt| \Xvec_{\rm (G)}^{\hat{\mu}} \lt(\tau + \Dl \tau, \xvec + \Dl \xvec\rt) \rt\rangle
- {i \over \hb} \, \Dl \tau \, (\Lie)^{\hat{\nu}} \, \Pvec_{\hat{\nu}} \,
| \Xvec_{\rm (G)}^{\hat{\mu}} \lt(\tau, \xvec\rt) \rangle,
\label{Lie-transport-ket}
\ee
where $(\Lie)^{\hat{\mu}} = \xi^{\hat{\nu}} (\nabla_{\hat{\nu}} \Xvec^{\hat{\mu}}) - \Xvec^{\hat{\nu}}
(\nabla_{\hat{\nu}} \xivec^{\hat{\mu}})$ is the Lie derivative along some vector field $\xivec^{\hat{\al}}$ tangent
to the reference worldline, as shown in Figure~\ref{fig:ket-rotation}, $\lt| \Xvec_{\rm (G)}^{\hat{\mu}}
\lt(\tau + \Dl \tau, \xvec + \Dl \xvec\rt) \rt\rangle
= \lt| \lt[\bar{e}^{\hat{\mu}}{}_\nu (x^\al + \Dl x^\al) \rt] (x^\nu + \Dl x^\nu) \rt\rangle$,
and $ \Pvec^{\hat{\al}}$ is the canonical momentum operator in local Lorentz frame co-ordinates.
To demonstrate path integral invariance under reparametrization of $\tau$, allow for
$(\Lie)^{\hat{\mu}} \rightarrow V^0 \, (\Lie)^{\hat{\mu}}$, where $V^0$ is the time component of the
four-vector $V^\mu = \Dl x^\mu/\Dl \tau$, and serves as the lapse function.
Obviously, $\tau$ is the proper time when $V^0 = 1$.

For what follows, assume that all indices are unhatted and refer to local Lorentz frame co-ordinates.
By making use of (\ref{U-Proj.}), it follows that
\be
\lt| \Xvec_{\rm (G)}^{'\mu} \lt(\tau + \Dl \tau, \xvec + \Dl \xvec\rt) \rt\rangle
& = & U_{\Dl \tau} (V^\al, \xivec^\al) \, | \Xvec_{\rm (G)}^\mu \lt(\tau, \xvec\rt) \rangle,
\label{time-translation}
\ee
where
\be
U_{\Dl \tau} (V^\al, \xivec^\al)
& = & 1 - {i \over \hb} \, V^0 \, \Dl \tau \lt\{ \lt[\dl^0{}_\al - {\cal F}(\tilde{R})_{\al \bt} \, {V^\bt \over V^0}
+ (\Lie)_\al \rt] \rt.
\nn
&  &{} + \lt. {i \over \hb} \, \eta_{\al \bt} \, \tilde{R}_{\lm \sg} \, x^\sg \,  {V^\bt \over V^0} \, \Pvec^\lm
\rt\} \Pvec^\al
\label{U-time-trans.}
\ee
is the infinitesimal local time translation operator, and
\be
{\cal F}(\tilde{R})_{\al \bt} & = & \lt(\tilde{R}_{\al \bt,\mu}
+ {1 \over 2!} \, \tilde{R}_{\al \sg, \bt \mu} \, x^\sg\rt) \Dl x^\mu
\nn
&  &{}
+ \lt({1 \over 2!} \, \tilde{R}_{\al \bt, \mu \nu}
+ {1 \over 3!} \, \tilde{R}_{\al \sg, \bt \mu \nu} \, x^\sg\rt) \Dl x^\mu \, \Dl x^\nu + O\lt((\Dl x)^3\rt).
\label{F(R)=}
\ee
It should be noted that
\be
U_{\Dl \tau}^{-1} (V^\al, \xivec^\al) & = & U_{-\Dl \tau} (V^\al, \xivec^\al)
\ \neq \ U_{\Dl \tau}^\dag (V^\al, \xivec^\al).
\label{U-time-trans.-property}
\ee
The fact that (\ref{U-time-trans.}) does not satisfy unitarity is not particularly surprising, since it is well-known
from spin-1/2 particle quantum mechanics in curved space-time that the respective Dirac Hamiltonian is not strictly
Hermitian \cite{Parker,Singh-Mobed-zitt}.
At the same time, the explicit coupling of $\hbar$ with space-time curvature, as found in
the last term in (\ref{U-time-trans.}),
has unexpected and important implications, namely the demonstration of quantum mechanical
weak equivalence principle violation, as shown later in this Letter.

\section{Configuration Space Path Integral in Curved Space-Time}

\subsection{Formal Expression for the Path Integral}

Having now obtained (\ref{U-time-trans.}),
it is relatively straightforward to determine the configuration space path integral, though with some subtle
new features.
Adopting the Heisenberg representation \cite{Sakurai} for the initial and final position ket vectors
$\lt|{\Xvec}_{\rm (i)}^{\rm (G)} (\tau_{\rm i}, \xvec_{\rm i}) \rt\rangle$ and
$\lt|{\Xvec'}_{\rm (f)}^{\rm (G)} (\tau_{\rm f}, \xvec_{\rm f}) \rt\rangle =
U_{(\tau_{\rm f} - \tau_{\rm i})}^{-1} (V^\al, \xivec^\al)
\lt|{\Xvec}_{\rm (f)}^{\rm (G)} (\tau_{\rm i}, \xvec_{\rm i}) \rt\rangle$,
let $\tau_{\rm f} - \tau_{\rm i} = N \Dl \tau$ for integer $N$, such that the scalar particle propagator can be
written as
\be
\lefteqn{\lt\langle \Xvec_{\rm (f)}^{'\rm (G)} (\tau_{\rm f}, \xvec_{\rm f}) \rt. \Big| \lt.
\Xvec_{\rm (i)}^{\rm (G)} (\tau_{\rm i}, \xvec_{\rm i}) \rt\rangle \ = \
\lt\langle \Xvec_{\rm (f)}^{\rm (G)} (\tau_{\rm i}, \xvec_{\rm i}) \rt.
\Big| U_{N \Dl \tau}^{-1 \dag} \Big| \lt.
\Xvec_{\rm (i)}^{\rm (G)} (\tau_{\rm i}, \xvec_{\rm i}) \rt\rangle }
\nn
&&{} = \
\lt\langle \Xvec_{\rm (f)}^{\rm (G)} (\tau_{\rm i}, \xvec_{\rm i}) \rt.
\Big| \lt(\prod_{n=1}^N \vec{1}_{(n)} \, U_{\Dl \tau}^{-1 \dag} \rt) \vec{1}_{(0)}
\Big| \lt.{\Xvec}_{\rm (i)}^{\rm (G)} (\tau_{\rm i}, \xvec_{\rm i}) \rt\rangle,
\label{propagator-formal}
\ee
where
\be
\vec{1}_{(n)} & = & \int_{-\infty}^\infty \d^3 \Xvec_{(n)}
\lt| \Xvec_{(n)} (\tau_n, \xvec_n) \rt\rangle \
\lt\langle \Xvec_{(n)} (\tau_n, \xvec_n) \rt|
\label{completeness}
\ee
is defined in terms of strictly flat space-time position ket vectors.
The two extra identity operators $\vec{1}_{(0)}$ and $\vec{1}_{(N)}$ are inserted in order to determine that,
to leading order in curvature,
\be
\lt\langle \Xvec_{(0)} \Big|
\Xvec_{\rm (i)}^{\rm (G)} (\tau_{\rm i}, \xvec_{\rm i}) \rt\rangle
& = & \int_{-\infty}^\infty {\d^3 \Pvec_{(0)} \over (2 \pi \hb)^3}
\lt[1 + {i \over \hb} \, \tilde{R}_{ij} (\tau_{\rm i}, \xvec_{\rm i}) \Xvec^j_{\rm (i)} \, \Pvec^i_{(0)}\rt]
\nn
&  &{} \times \exp\lt[{i \over \hb} \lt(\Xvec_{(0)} - \Xvec_{{\rm (i)}}\rt) \cdot \Pvec_{(0)}\rt]
\nn
& \approx & \dl^3 \lt(\Xvec_{(0)} - \lt[\Xvec_{\rm (i)}
- \tilde{R}_{ij} (\tau_{\rm i}, \xvec_{\rm i}) \Xvec^j_{\rm (i)} \, \hat{\xvec}^i \rt]\rt),
\label{<X0|XG>=}
\nl
\lt\langle \Xvec_{\rm (f)}^{\rm (G)} (\tau_{\rm i}, \xvec_{\rm i}) \Big|
\Xvec_{(N)} \rt\rangle
& \approx & \dl^3 \lt(\Xvec_{(N)} - \lt[\Xvec_{\rm (f)}
- \tilde{R}_{ij} (\tau_{\rm i}, \xvec_{\rm i}) \Xvec^j_{\rm (f)} \, \hat{\xvec}^i \rt]\rt).
\label{<XG|XN>=}
\ee

Given that the Hamiltonian for a free scalar particle is $H(\Pvec) = \sqrt{\Pvec \cdot \Pvec + m^2}$,
it is possible to integrate out the intermediate momentum states exactly to first-order in
$\tilde{R}_{\mu \nu}$.
This is accomplished by employing the integral representation of the modified Bessel function
\cite{Ferraro,Gradshteyn}
\be
K_{\pm \nu} \lt(\mu \bt\rt) & = & {\bt^{-\nu} \over 2} \, e^{-i \nu \pi/2} \int_0^\infty \d N \, N^{\nu - 1} \,
\exp \lt[{i \, \mu \over 2} \lt(N - {\bt^2 \over N}\rt)\rt] \, ,
\label{modified-Bessel-integral}
\ee
for $\nu = 1/2$, with ${\rm Im}(\mu) > 0$ and ${\rm Im}(\mu \bt^2) < 0$.
Assuming that ${\rm Im}(\Dl \tau) \lesssim 0$ and identifying
\be
\mu & = & -{V^0 \, \Dl \tau \over \lmbar} \, ,
\label{mu=}
\nl
\bt & = &
-{i \over m} \lt[1 - {\cal G}(\tilde{R})_{0\al} \, {V^\al \over V^0} + (\Lie)_0
+ {i \over \hb} \lt(\tilde{R}_{0\al} \, x^\al \, {V_j \over V^0} - \tilde{R}_{j\al} \, x^\al\rt) \Pvec^j \rt]
\nn
&  &{} \times \sqrt{\Pvec \cdot \Pvec + m^2} \, ,
\label{bt=}
\ee
where
\be
{\cal G}(\tilde{R})_{\mu \nu} & = & {\cal F}(\tilde{R})_{\mu \nu} + \tilde{R}_{\mu \nu}
+ \tilde{R}_{\mu \al, \nu} \, x^\al + (\tilde{R}^\al{}_\al) \, \eta_{\mu \nu} \, ,
\label{G(R)=}
\ee
and $\lmbar = \hb/m$ is the scalar particle's reduced Compton wavelength,
(\ref{propagator-formal}) can be recast into Gaussian form with respect to integration over
$\Pvec_{(n)}$ at local time $\tau_{(n)}$.

Much of the relevant details for computing the free-particle scalar propagator
are deferred to the forthcoming longer publication \cite{Singh-Mobed-prep-1}.
However, it is relatively straightforward to demonstrate that
when $V^0 \Dl \tau \rightarrow \d \tau$, $(V^0 \Dl \tau)^{-1} \rightarrow \dl(0)$,
and $V^\mu/V^0 \rightarrow \dot{x}^\mu(\tau) = (1, \vec{\dot{x}}(\tau))$, the integration measure
for the skeletonized path integral becomes
\be
\lim_{N \rightarrow \infty} \lt(1 \over 2 \pi \lmbar \, i \, V^0 \Dl \tau\rt)^{3N/2} \prod_{n=0}^N
\int_{-\infty}^\infty \d^3 \Xvec_{(n)} & \rightarrow & \int {\cal D} \lt[\Xvec(\tau)\rt] \, ,
\label{integration-measure}
\ee
and the configuration space propagator in curved space-time becomes
\be
\lefteqn{
\lt\langle \Xvec_{\rm (f)}^{'\rm (G)} (\tau_{\rm f}, \xvec_{\rm f}) \rt. \Big| \lt.
\Xvec_{\rm (i)}^{\rm (G)} (\tau_{\rm i}, \xvec_{\rm i}) \rt\rangle \ = \ \int {\cal D} \lt[\Xvec(\tau)\rt]
 \exp \lt[{i \over \hb} \int_{\tau_{\rm i}}^{\tau_{\rm f}}
\d \tau \, L^{\rm (G)}_{(\rm Re,0)} \rt]
}
\nn
&& \times \exp \lt[{i \over \hb} \int_{\tau_{\rm i}}^{\tau_{\rm f}}
\d \tau \lt(L^{\rm (G)}_{(\rm Re,1)}
+ i \lt[L^{\rm (G)}_{(\rm Im,0)} + \lmbar \, \dl(0) \, L^{\rm (G)}_{(\rm Im,1)} \rt] \rt) \rt] \, ,
\label{propagator-continuum}
\ee
where the free-particle Lagrangian $L^{\rm (G)}(\tau, \xvec(\tau), \vec{\dot{x}}(\tau))
= L^{\rm (G)}_{(\rm Re,0)} + L^{\rm (G)}_{(\rm Re,1)}
+ i \lt[L^{\rm (G)}_{(\rm Im,0)} + \lmbar \, \dl(0) \, L^{\rm (G)}_{(\rm Im,1)} \rt]$ is expressed in terms of
\be
L^{\rm (G)}_{(\rm Re,0)} & = & -m \lt[1 - {1 \over 2} \lt\{2 \, {\cal G}(\tilde{R})_{(00)}
+ 4 \, {\cal G}(\tilde{R})_{(0j)} \, \dot{x}^j + \lt[\eta_{ij} + 2 \, {\cal G}(\tilde{R})_{(ij)}\rt] \,
\dot{x}^i \, \dot{x}^j
\rt\} \rt]
\nn
& \approx & -m \lt[-g^{\rm (eff.)}_{\mu \nu} \, \dot{x}^\mu \, \dot{x}^\nu \rt]^{1/2} \, ,
\qquad g^{\rm (eff.)}_{\mu \nu} \ = \ \eta_{\mu \nu} + 2 \, {\cal G}(\tilde{R})_{(\mu \nu)} \, ,
\label{LG-Re,0=}
\ee
the expected free-particle Lagrangian in curved space-time that generates the geodesic equation, along with
\be
L^{\rm (G)}_{(\rm Re,1)} & = & -m \lt\{(\Lie)_\mu + 2 \lt[(\Lie)_0 \, \dl^0{}_\mu
- {\cal G}(\tilde{R})_{0\mu} \rt] \eta_{ij} \, \dot{x}^i \, \dot{x}^j\rt\} \dot{x}^\mu \, ,
\label{LG-Re,1=}
\nl \nn
L^{\rm (G)}_{(\rm Im,0)} & = & m \lt\{\lmbar \lt[(\Lie)^\al{}_{,\al} - {\cal H}(\tilde{R})_\al \, \dot{x}^\al\rt]
+ {1 \over 2 \lmbar} \, \tilde{R}_{k\al} \, x^\al \, \dot{x}^k \lt(1 - 3 \, \eta_{ij} \, \dot{x}^i \, \dot{x}^j\rt) \rt.
\nn
&  &{} - \lt. {1 \over \lmbar} \, \tilde{R}_{0\al} \, x^\al \lt(1 - 3 \, \eta_{ij} \, \dot{x}^i \, \dot{x}^j\rt)
\lt(1 - \eta_{kl} \, \dot{x}^k \, \dot{x}^l\rt) \rt\} \, ,
\label{LG-Im,0=}
\nl \nn
L^{\rm (G)}_{(\rm Im,1)} & = & m \lt\{{3 \over 2} \lt[(\Lie)_0 - {\cal G}(\tilde{R})_{00}\rt]
+ {5 \over 2} \lt[(\Lie)_j - {\cal G}(\tilde{R})_{j0}\rt] \dot{x}^j \rt.
\nn
&  &{} - \lt. {5 \over 4} \lt[\eta_{ij} + 2 \, {\cal G}(\tilde{R})_{(ij)}\rt] \dot{x}^i \, \dot{x}^j
+ \lt[(\Lie)_0 - {\cal G}(\tilde{R})_{0\mu} \, \dot{x}^\mu \rt] \eta_{ij} \, \dot{x}^i \, \dot{x}^j \rt\} \, ,
\nn
\label{LG-Im,1=}
\ee
where
\be
{\cal H}(\tilde{R})_\mu & = & {\cal F}(\tilde{R})^\al{}_{\mu,\al} + \tilde{R}^\al{}_{\al,\mu} \, ,
\label{H(R)=}
\ee
and
\be
\Xvec^{\rm (G)}(\tau_{\rm (i,f)}) = \Xvec(\tau_{\rm (i,f)}) - \tilde{R}_{ij}(\tau_{\rm i}, \xvec_{\rm i})
\, \Xvec^j(\tau_{\rm (i,f)}) \, \hat{\xvec}^i
\label{Xi,f}
\ee
from (\ref{<X0|XG>=}) and (\ref{<XG|XN>=}) in the continuum limit.

\subsection{Physical Consequences}

Even at a purely formal level, there are some valuable insights to be gained
from a preliminary analysis of (\ref{propagator-continuum}).
First, it contains the correct free-particle Lagrangian (\ref{LG-Re,0=}) in the weak-field
limit, with additional terms in (\ref{LG-Re,1=}) that both preserve the conservation
of probability and satisfy the weak equivalence principle.
In contrast, all the probability violating contributions due to curvature, namely
(\ref{LG-Im,0=}) and (\ref{LG-Im,1=}), show a direct coupling of $\lmbar$ 
with the gravitational background, yielding a {\em quantum violation} of the
weak equivalence principle at the Compton wavelength scale.

The fact that {\em only} the probability non-conserving terms in the scalar propagator are responsible for
weak equivalence principle violation is especially interesting and deserves careful analysis.
In particular, it is straightforward to see that, under the interchange of $\tau_{\rm i} \leftrightarrow \tau_{\rm f}$
and primed-to-unprimed ket vector notation in (\ref{propagator-continuum}) followed by taking its Hermitian conjugate,
the transformed scalar propagator now reveals the existence of a time reversal symmetry breakdown,
precisely due to (\ref{LG-Im,0=}) and (\ref{LG-Im,1=}).
This observation may provide a potentially satisfactory reason for why there exists a macroscopic
preference for time to propagate in the forward direction only.
Given that there is already some potential evidence for weak equivalence principle violation at the
Compton wavelength scale in the context of spin-1/2 particle propagation in curved space-time
\cite{Singh-Mobed-zitt}, this particular observation gained from
(\ref{propagator-continuum}) is worthy of future exploration.

\section{Regularization of the Path Integral}

\subsection{Cartesian Co-ordinates}

Explicit evaluation of (\ref{propagator-continuum}) in skeletonized form is
a straightforward exercise involving multiple Gaussian integrations with
respect to $\prod_{n=1}^{N-1} \d^3 \Xvec_{(n)}$.
However, it is necessary to then regularize the propagator in order to remove all
of its singular contributions in the limit as $V^0 \rightarrow 0$.
Normally, this involves describing (\ref{propagator-continuum}) according to the ansatz
\be
\lefteqn{\lt\langle \Xvec_{\rm (f)}^{'\rm (G)} (\tau_{\rm f}, \xvec_{\rm f}) \rt. \Big| \lt.
\Xvec_{\rm (i)}^{\rm (G)} (\tau_{\rm i}, \xvec_{\rm i}) \rt\rangle_{\rm (reg.)}
\ \equiv \ \lt(1 \over 2 \pi \lmbar \, i \, V^0 \lt(\tau_{\rm f} - \tau_{\rm i}\rt) \rt)^{3/2}}
\nn
&& \times \ \exp \lt[{i \over 2} \, {m \over \hb} \,
{\lt(\eta_{\mu \nu} \, \Dl \Xvec^\mu_{\rm(i \rightarrow f)} \, \Dl \Xvec^\nu_{\rm(i \rightarrow f)} \rt)
\over  V^0 \lt(\tau_{\rm f} - \tau_{\rm i}\rt)}
\rt]
\sum_{k = 0}^\infty a_k(x^\mu_{\rm i}, x^\mu_{\rm f}) \lt(i \, V^0\rt)^k \, ,
\label{propagator-Seeley-DeWitt}
\ee
where $\Dl \Xvec^\mu_{\rm(i \rightarrow f)} = \Xvec^\mu_{\rm (f)} - \Xvec^\mu_{\rm (i)}$
and $a_k(x^\mu_{\rm i}, x^\mu_{\rm f})$ are the curvature-dependent Seeley-DeWitt coefficients
\cite{Kleinert,Parker-Toms,Fulling}, whose values are determined from solving the heat kernel equation.
For (\ref{propagator-continuum}) in skeletonized form, it so happens that it is unnecessary to
perform this calculation to achieve this objective, since the propagator prior to the Gaussian integrations
can easily be put into power series form consistent with (\ref{propagator-Seeley-DeWitt}),
such that all inverse powers of $V^0$ are identified by inspection alone
and subsequently removed by hand.

While the relevant details of this computation are deferred to the aforementioned longer publication
\cite{Singh-Mobed-prep-1}, it follows that the regularized scalar particle propagator is
\be
\lefteqn{\lt\langle \Xvec_{\rm (f)}^{'\rm (G)} (\tau_{\rm f}, \xvec_{\rm f}) \rt. \Big| \lt.
\Xvec_{\rm (i)}^{\rm (G)} (\tau_{\rm i}, \xvec_{\rm i}) \rt\rangle_{\rm (reg.)}
 \ = \ }
\nn
&& \lim_{N \rightarrow \infty}
\lt(1 \over 2 \pi \lmbar \, i \, V^0 \lt(\tau_{\rm f} - \tau_{\rm i}\rt) \rt)^{3/2}
\exp \lt[{i \over 2} \, {m \over \hb} \,
{\lt(\eta_{\mu \nu} \, \Dl \Xvec^\mu_{\rm(i \rightarrow f)} \, \Dl \Xvec^\nu_{\rm(i \rightarrow f)} \rt)
\over V^0 \lt(\tau_{\rm f} - \tau_{\rm i}\rt)} \rt]
\nn \nn
&&{} \times \sum_{k = 0}^\infty \sum_{n = 1}^N {1 \over k!} \lt\{
\lt[C^{(k,0)}_{(n)}(x^\mu)
+ \sum_{l=1}^4 {k! \over (k+l)!} \lt(-i \over 2N\rt)^l \, {C^{(k+l,l)}_{(n)}(x^\mu) \over \lmbar^l} \rt] \rt.
\nn \nn
&&{} + \lt.
\exp\lt[{n \over N} \, \Dl \Xvec^\al_{\rm(i \rightarrow f)} \, {\partial \over \partial x^\al} \rt]
\lt[C^{(k,0)}_{({\cal L},n)}(x^\mu)
+ \sum_{l=1}^2 {k! \over (k+l)!} \lt(-i \over 2N\rt)^l \, {C^{(k+l,l)}_{({\cal L},n)}(x^\mu) \over \lmbar^l} \rt]
\rt\}
\nn \nn
&&{} \times
\lt(-i \, V^0 \, \Dl \tau \over 2 \lmbar\rt)^k
\exp \lt[-i \, {m \over \hb} \lt(
\tilde{R}_{(ij)} (\tau_{\rm i}, \xvec_{\rm i}) \, \Dl \Xvec^i_{\rm(i \rightarrow f)} \,
\Dl \Xvec^j_{\rm(i \rightarrow f)}
\over V^0 \lt(\tau_{\rm f} - \tau_{\rm i}\rt)\rt)\rt] \, ,
\label{propagator-skeletonized}
\ee
where $N \Dl \tau = \tau_{\rm f} - \tau_{\rm i} \equiv 1$ and, to leading-order in curvature,
the coefficients $C^{(k,l)}_{(n)}(x^\mu)$ and $C^{(k,l)}_{({\cal L},n)}(x^\mu)$---the latter of which are
proportional to $(\Lie)^\mu(\tau_{\rm i}, \xvec_{\rm i})$---are generally {\em complex-valued}.
Again, only the probability non-conserving imaginary parts denote a quantum violation of the
weak equivalence principle via the explicit coupling of $m$ to the gravitational background,
as evidenced in the case of $l = 0$, where
\be
C^{(k,0)}_{(n)}(x^\mu) & = & 1 - \lt(2k - {3 \over 2}\rt) {\cal G}(\tilde{R})_{00}
+ 2 i \, k \lt(\lmbar \, {\cal H}(\tilde{R})_0 + {1 \over \lmbar} \, \tilde{R}_{00} \, \tau_{(n)}
\rt.
\nn
&  &{} + \lt.
{1 \over \lmbar} \, \tilde{R}_{0j} \lt[\Xvec^j_{\rm (i)} + {n \over N} \, \Dl \Xvec^j_{\rm(i \rightarrow f)}
\rt] \rt) \, ,
\label{C-n}
\nl \nn
C^{(k,0)}_{({\cal L},n)}(x^\mu) & = & \lt(2k - {3 \over 2}\rt) (\Lie)_0 (\tau_{\rm i}, \xvec_{\rm i})
- 2 i \, k \, \lmbar \, (\Lie)^\al{}_{,\al} (\tau_{\rm i}, \xvec_{\rm i}) \, .
\label{C-n-Lie}
\ee

In addition to these new physical predictions, there also emerges the prediction of a
curvature-dependent and {\em gauge-invariant} overall phase factor in the propagator,
as found in the final line of (\ref{propagator-skeletonized}), that is dependent upon the
spatial separation between the initial and final position states.
This phase factor can be interpreted as a manifestation of the gravitational Aharonov-Bohm effect
\cite{Aharonov-Bohm,Dowker,Ford,Bezerra,Heller} and Berry's phase
\cite{Berry's-Phase,Singh-Mobed-Papini,Cai-Papini}, each of which are the subjects of extensive study already.
Such an effect to come from the path integral should reveal important consequences that
are also worthy of future consideration.

\subsection{Curvilinear Co-ordinates}

It is also straightforward to evaluate (\ref{propagator-continuum}) in terms of curvilinear
co-ordinates, following the established techniques first put forward by Kleinert \cite{Kleinert,Kleinert-1}.
This is done by letting $\Xvec^\mu (\tau, \xvec) = \Xvec^\mu (\tau, \xvec(q))$, where $q$ are the generalized
spatial curvilinear co-ordinates.
Following Kleinert's treatment of the integration measure $\prod_{n=1}^{N-1} \d^3 \Xvec_{(n)} \rightarrow
\prod_{n=1}^{N-1} \d^3 (\Dl \Xvec_{(n)})$ to integrate over $\Dl \Xvec^j_{(n)}$ in terms of the pre-point
description of
$\Xvec^j_{(n)} = \Xvec^j_{(n-1)} + \Dl \Xvec^j_{(n)} = \Xvec^j_{(0)} + \sum_{k=1}^n \Dl \Xvec^j_{(k)}$,
it follows that $\d^3 (\Dl \Xvec_{(n)}) \rightarrow \d^3 (\Dl q_{(n)}) \, J(q_{(n)})$, where
$J(q_{(n)})$ is the Jacobian expressed in terms of the affine connection
$\hat{\Gm}^i{}_{jk} (\tau_{(n)})$ which satisfies the geodesic equation \cite{Kleinert,Kleinert-1}
\be
{\d^2 q^i_{(n)} \over \d \tau^2} + \hat{\Gm}^i{}_{jk} (\tau_{(n)}) \, {\d q^j_{(n)} \over \d \tau}
\, {\d q^k_{(n)} \over \d \tau} & = & 0 \,
\label{geodesic-equation-curvilinear}
\ee
for spatial curvature in the absence of torsion.
This ensures that the short-time action to connect infinitesimally separated events in space-time
describes classical motion in the presence of a locally curved background.

Again, with more details to be shown in the longer publication \cite{Singh-Mobed-prep-1}, it is determined
that the regularized scalar particle propagator in curvilinear co-ordinates takes the form
\be
\lefteqn{\lt\langle \Xvec_{\rm (f)}^{'\rm (G)} (\tau_{\rm f}, \xvec_{\rm f}) \rt. \Big| \lt.
\Xvec_{\rm (i)}^{\rm (G)} (\tau_{\rm i}, \xvec_{\rm i}) \rt\rangle_{\rm (reg.)}
 \ = \ }
\nn
&& \lim_{N \rightarrow \infty} \exp \lt[-{i \over 2} \, {N \, \Dl \tau \over \lmbar \, V^0}\rt]
\exp \lt[\lt(i \, \lmbar \, V^0 \, \Dl \tau\rt) \sum_{s=1}^{N-1} {\cal N}_{(0)}(\hat{\Gm}_{(s)})\rt]
\nn
&&{}
\times \lt(1 \over 2 \pi \lmbar \, i \, V^0 \, \Dl \tau \rt)^{3/2}
\exp \lt[{i \over 2} \, {m \over \hb} \,
{\lt(\eta_{ij} \, \Dl \Xvec^\mu_{(N)} \, \Dl \Xvec^\nu_{(N)} \rt)
\over V^0 \, \Dl \tau} \rt]
\nn \nn
&&{} \times \sum_{k = 0}^\infty \sum_{n = 1}^N {1 \over k!} \lt\{
\lt[\lt\langle C^{(k,0)}_{(n)}(x^\mu)\rt\rangle
+ \sum_{l=1}^4 {k! \over (k+l)!} \lt(-i \over 2N\rt)^l \, {\lt\langle C^{(k+l,l)}_{(n)}(x^\mu)\rt\rangle
\over \lmbar^l} \rt] \rt.
\nn \nn \nn
&&{} + \exp\lt[N \, \Dl \tau \, {\partial \over \partial \tau} \rt] \exp \lt[\Dl \Xvec^j_{(N)} \,
{\partial \over \partial x^j} \rt]
\exp\lt[\lt(i \, \lmbar \, V^0 \, \Dl \tau\rt) \lt\{{1 \over 2} \, \bar{h}^{ab} \,
{\partial \over \partial q^a} \, {\partial \over \partial q^b} \rt. \rt.
\nn \nn
&&{} +  \sum_{s=1}^{N-1} \lt[{\cal N}^a_{({\cal L},1)}(\hat{\Gm}_{(s)}) \,
{\partial \over \partial q^a} + \lt(i \, \lmbar \, V^0 \, \Dl \tau\rt)
{\cal N}^{ab}_{({\cal L},2)}(\hat{\Gm}_{(s)}) \, {\partial \over \partial q^a} \, {\partial \over \partial q^b}
\rt.
\nn \nn
&&{} + \lt. \lt. \lt. \lt(i \, \lmbar \, V^0 \, \Dl \tau\rt)^2
{\cal N}^{abcd}_{({\cal L},4)}(\hat{\Gm}_{(s)}) \, {\partial \over \partial q^a} \, {\partial \over \partial q^b}
\, {\partial \over \partial q^c} \, {\partial \over \partial q^d} \rt] \rt\} \rt]
\, \lt[\lt\langle C^{(k,0)}_{({\cal L},n)}(x^\mu)\rt\rangle \rt.
\nn \nn
&&{} + \lt. \lt. \sum_{l=1}^2 {k! \over (k+l)!} \lt(-i \over 2N\rt)^l \,
{\lt\langle C^{(k+l,l)}_{({\cal L},n)}(x^\mu) \rt\rangle \over \lmbar^l} \rt]
\rt\} \lt(-i \, V^0 \, \Dl \tau \over 2 \lmbar\rt)^k \, ,
\label{propagator-skeletonized-curvilinear}
\ee
where $\bar{h}^{ab} = \eta^{ij} \, (\partial q^a/\partial x^i) \, (\partial q^b/\partial x^j)$
is the inverse of the induced metric $h_{ab} = \eta_{ij} \, (\partial x^i/\partial q^a) \,
(\partial x^j/\partial q^b)$ in curvilinear co-ordinates,
${\cal N}^{abc\cdots}(\hat{\Gm})$ is a multi-indexed function of the affine connection, and
$\lt\langle C^{(k,l)} (x^\mu)\rt\rangle$ in (\ref{propagator-skeletonized-curvilinear})
comprises of terms coupled to even powers of $\Dl \Xvec^j_{(n)}$ that survive the integration
over $\d^3 (\Dl \Xvec_{(n)})$, with $\Xvec_{(0,N)} = \Xvec^{\rm (G)}(\tau_{\rm (i,f)})$ from (\ref{Xi,f}).

\section{Conclusion}

This Letter demonstrates a fundamentally different approach to path integral quantum mechanics
in curved space-time for a scalar relativistic particle, strongly motivated by Feynman's
original vision of a sum-over-histories, but in terms of propagation due to a non-unitary local time
translation operator that incorporates local curvature contributions.
It reveals what appear to be significant physical predictions, with potentially broad
implications concerning quantum mechanical interactions in a non-trivial gravitational field.

From this approach shown above, a configuration space propagator for a particle of mass $m$ is derived
that reveals the prediction of a quantum violation of the weak equivalence principle
at the Compton wavelength scale, some {\em twenty orders of magnitude larger} than the
Planck scale if $m$ is a subatomic particle.
This indicates the significant potential for identifying fundamental relationships
between quantum matter and space-time curvature that would otherwise be obscured from view
by adopting a strictly unitary time translation perspective.
The fact that {\em only} the probability non-conserving contributions
to the propagator violate the weak equivalence principle suggests the prediction
that time reversal symmetry is satisfied {\em only} at a strictly classical length scale,
and that its breakdown results in a preferred arrow of time motivated by
quantum mechanics alone.
Furthermore, there exists the prediction of a curvature-dependent and gauge-invariant phase factor
identified as the gravitational analogue of the Aharonov-Bohm effect and Berry's phase,
worthy of a deeper investigation to probe its physical consequences.

It is worthwhile to consider further developments of this approach when applied to non-zero integer
and half-integer spin particles \cite{Singh-Mobed-prep-2}.
The results derived from this treatment may provide the foundation for later considering a
many-body particle treatment in curved space-time, ultimately leading towards an alternative approach to
quantum field theory in curved space-time that may be compared with established approaches contained
within the literature \cite{Parker-Toms}.

\section{Acknowledgements}

The authors thank Lewis Ryder for helpful comments and suggestions regarding this Letter.


\end{document}